\documentclass[a4paper,11pt]{article}
\usepackage{pos}
\usepackage{xspace}
\usepackage{hyperref}
\usepackage{wrapfig}
\usepackage{lineno}
\usepackage{relsize}
\usepackage{caption}
\usepackage{subcaption}
\newcommand{\dndeta}{\ensuremath{\mathrm{d}N_{\rm ch}/\mathrm{d}\eta}\xspace}
\newcommand{\avdndeta}{\ensuremath{\langle\dndeta\rangle}\xspace}
\newcommand{\npart}{\ensuremath{N_{\mathrm{part}}}\xspace}
\newcommand{\pp}{\ensuremath{\mathrm {p\kern-0.05em p}}\xspace}
\newcommand{\PbPb}{\ensuremath{\mbox{Pb--Pb}}\xspace}
\newcommand{\snn}{\ensuremath{\sqrt{s_{\mathrm{NN}}}}\xspace}
\newcommand{\s}{\ensuremath{\sqrt{s}}\xspace}

\title{Charged-particle production in \pp collisions at \s = 13.6 TeV and \PbPb collisions at \snn = 5.36 TeV with ALICE}
\ShortTitle{\dndeta in \pp and \PbPb collisions}

\manuallySeparateAuthors

\author*[a]{Abhi Modak}
\author{ for the ALICE Collaboration}

\affiliation[a]{University of Brescia, Italy}

\emailAdd{abhi.modak@cern.ch}

\abstract{This article presents the measurement of charged-particle pseudorapidity ($\eta$) density, \dndeta, in proton--proton (\pp) collisions at a centre-of-mass energy \s = 13.6 TeV, and in lead--lead (\PbPb) collisions at a centre-of-mass energy per nucleon pair \snn = 5.36 TeV. The analysis is performed using the Run~3 data recorded during 2022 and 2023 by the upgraded ALICE detector. The charged-particle multiplicity is measured at midrapidity ($|\eta|<1$) using the new monolithic active pixel sensors-based Inner Tracking System and the Time Projection Chamber upgraded with Gas Electron Multiplier-based readout system. The measurements in \pp collisions are reported for inelastic events with at least one charged particle having $|\eta|<1$ whereas for \PbPb collisions, the \dndeta is obtained for different centrality classes, ranging from 0--5\% (most central) to 70--80\% (most peripheral). The energy dependence of average charged-particle pseudorapidity density (\avdndeta) measured in $|\eta|<0.5$ is studied and compared to earlier measurements at lower collision energies. In \PbPb collisions, the evolution of \avdndeta as a function of the average number of participating nucleons, $\langle \npart \rangle$, determined with a Glauber model, is also studied and compared with predictions from theoretical models.}

\FullConference{42nd International Conference on High Energy Physics (ICHEP2024)\\
18-24 July 2024\\
Prague, Czech Republic\\}

\begin{document}
\maketitle

\section{Introduction}
The pseudorapidity density of charged particles (\dndeta) produced in the central rapidity region is a key observable to investigate the underlying mechanisms of particle production and to characterise properties, such as the initial gluon and energy density, of the matter created in high-energy nuclear collisions. At collider energies, the production of final-state particles is driven by the interplay of soft and hard quantum chromodynamics (QCD) interactions and is sensitive to non-linear QCD evolution in the initial state. The study of \dndeta and its dependence on colliding system, centre-of-mass energy, and collision geometry is essential to understand the relative contributions of these interactions to the final-state particle production. Moreover, measurement of this fundamental observable provides important constraints to model calculations based on different particle production mechanisms and initial conditions.

In this article, we report on the study of the charged-particle pseudorapidity density in \pp collisions at \s = 13.6 TeV and in \PbPb collisions at \snn = 5.36 TeV, utilising the data taken with the ALICE detector. The average charged-particle density, \avdndeta, is measured in the pseudorapidity interval $|\eta|\equiv|-\ln \tan(\theta/2)|<0.5$, where $\theta$ is the polar angle between the charged-particle direction and the beam axis ($z$). The data are compared to model predictions and previous measurements at lower centre-of-mass energies. Our result provides new insight into particle production mechanisms in \pp and \PbPb collisions at the highest ever centre-of-mass energies achieved at the LHC.


\section{Experimental setup}
During the LHC Long Shutdown 2 (2019-2022), the ALICE detector underwent a significant upgrade that has led to a new experimental setup which is capable of recording data of \PbPb collisions at rates up to 50 kHz~\cite{ALICE:upgrade}. The major upgrades include shifting to continuous data-taking mode, upgrading several sub-detectors, and developing a new online-offline (O$^2$) software framework for online and offline data processing, as well as physics analysis.

The main sub-detectors used in this analysis are the upgraded Inner Tracking System (ITS) and the Time Projection Chamber (TPC), and the newly installed Fast Interaction Trigger (FIT). The new ITS~\cite{ALICE:ITSupgrade} consists of seven cylindrical layers equipped with silicon monolithic active pixel sensors, providing better pointing resolution thanks to its reduced distance to the interaction point (IP) and better position resolution compared to the ITS used in Runs 1 and 2. The upgraded TPC~\cite{ALICE:TPCupgrade} employs new readout chambers based on Gas Electron Multiplier (GEM) foils that minimise the ion backflow and the resulting space charge effects in the TPC. This allows the TPC to operate at interaction rates up to 50 kHz for \PbPb collisions. In this analysis, the ITS and TPC detectors are used to measure charged particles at midrapidity. The FIT detector~\cite{ALICE:FIT} consists of five distinct detector stations grouped into three sub-detectors (FT0, FV0, FDD) installed on both sides (A and C side) of the IP. This analysis uses FT0 for triggering, event selection, and centrality determination.


\section{Event selection and data analysis}
A set of event selection conditions have been considered in this analysis. Events are selected based on their timing relative to the bunch crossing and the corresponding FT0 signal. Events which are close in time to the ITS readout time frame borders are not considered. Events where the vertex position estimated from the FT0 is more than 1 cm away from the vertex position determined from TPC and ITS tracks are also removed from the analysis. The selected events are further required to have a reconstructed vertex within $\pm$ 10 cm from the nominal IP along the $z$ axis.

\begin{wrapfigure}{r}{0.48\textwidth}
  \vspace*{-0.8cm}
  \begin{center}
    \includegraphics[scale=0.38]{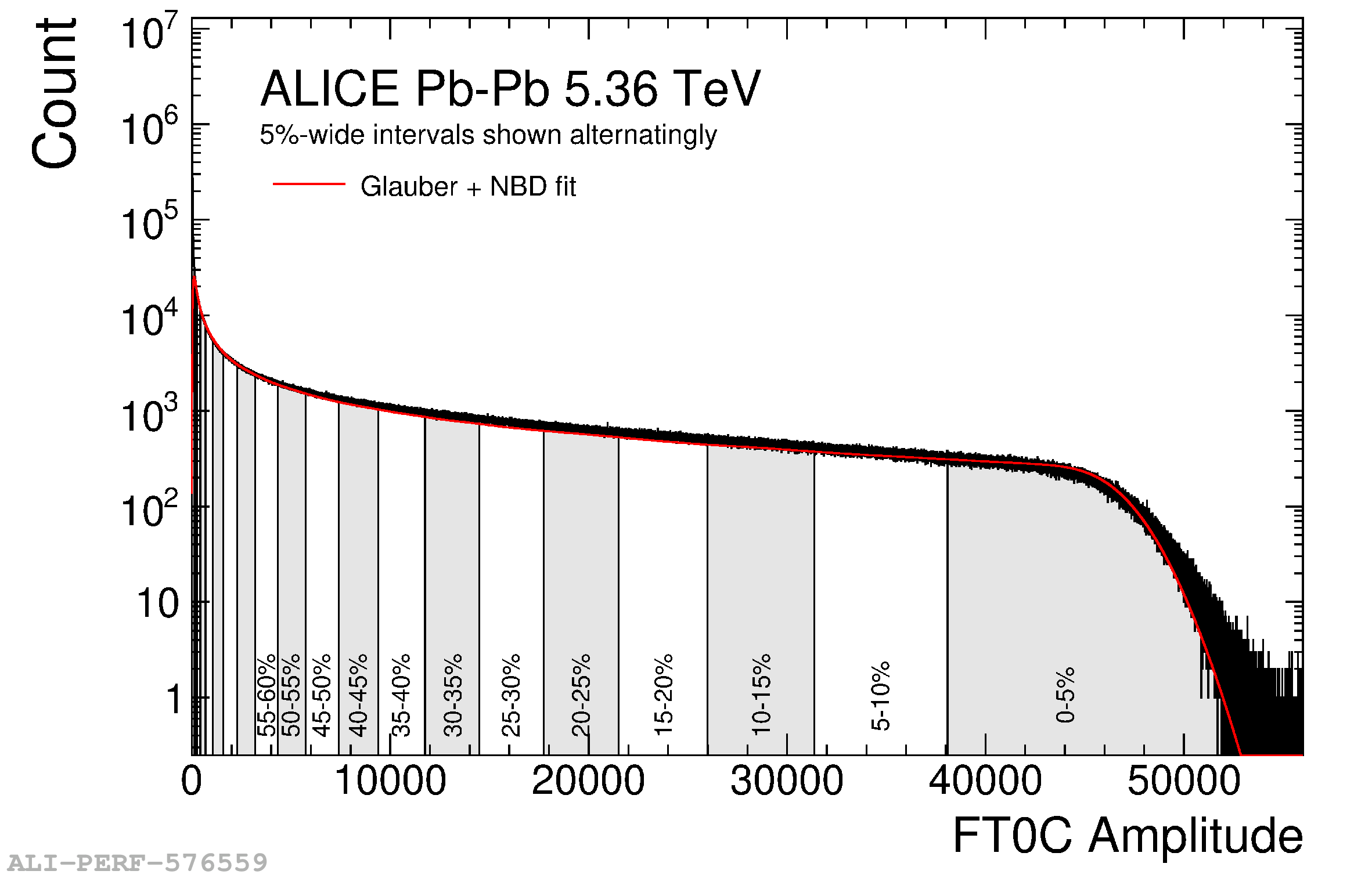}
  \end{center}
  \vspace*{-0.5cm}
  \caption{Distribution of the FT0C amplitude measured in \PbPb collisions at \snn = 5.36 TeV. The distribution is fitted with the Glauber MC combined with a NBD, shown as a red line. Centrality classes are also indicated by vertical lines.}
  \label{FT0C_cent}
\end{wrapfigure}

Centrality classes in \PbPb collisions are determined by fitting the measured FT0 amplitude on the C side, $-3.3<\eta<-2.1$, (see Fig.~\ref{FT0C_cent}) with a Glauber Monte Carlo (MC) model~\cite{Miller:2007ri} coupled with a negative binomial distribution (NBD), following the method developed previously~\cite{ALICE:2013hur}. The average number of participating nucleons in a given centrality class, $\langle \npart \rangle$, reflects the collision geometry and is obtained using the Glauber calculation by classifying events according to the impact parameter.

The charged-particle tracks for the analysis are categorised into \textit{global} tracks, \textit{ITS-only} tracks, and \textit{TPC-only} tracks. \textit{Global} tracks are the best quality tracks that are matched between ITS and TPC. The \textit{TPC-only} tracks are excluded from the analysis due to insufficient pointing resolution when propagated to the vertex and significant secondary particle contamination. The tracks considered in this study include \textit{global} tracks requiring to pass quality cuts for ITS and TPC contributions, and \textit{ITS-only} tracks passing quality cuts for ITS contributions. Both types of selected tracks are further required to satisfy a criterion on distance of closest approach to ensure they originate from primary particles.

The raw \dndeta obtained from the number of selected tracks within $|\eta|<1$ is corrected for the detector acceptance and the efficiency of a primary particle to produce a track using \textsc{Pythia} 8/Angantyr~\cite{Bierlich:2018xfw} MC event generator, following the method described in Ref.~\cite{ALICE:2015olq}. Systematic uncertainties from various sources (extrapolation to zero transverse momentum, particle composition, variations in detector acceptance, uncertainty on the centrality determination, variations in track selection) are evaluated and then added in quadrature to obtain the total systematic uncertainty. For \pp, the total systematic uncertainty amounts to $\sim$2\%, while for \PbPb, the total systematic uncertainty is 2.6\% (6.2\%) for the most central (most peripheral) collisions. The statistical uncertainties are found to be negligible.

\begin{figure}[h!]
  \centering
  \begin{minipage}[c]{0.4\textwidth}
    \includegraphics[scale=0.3]{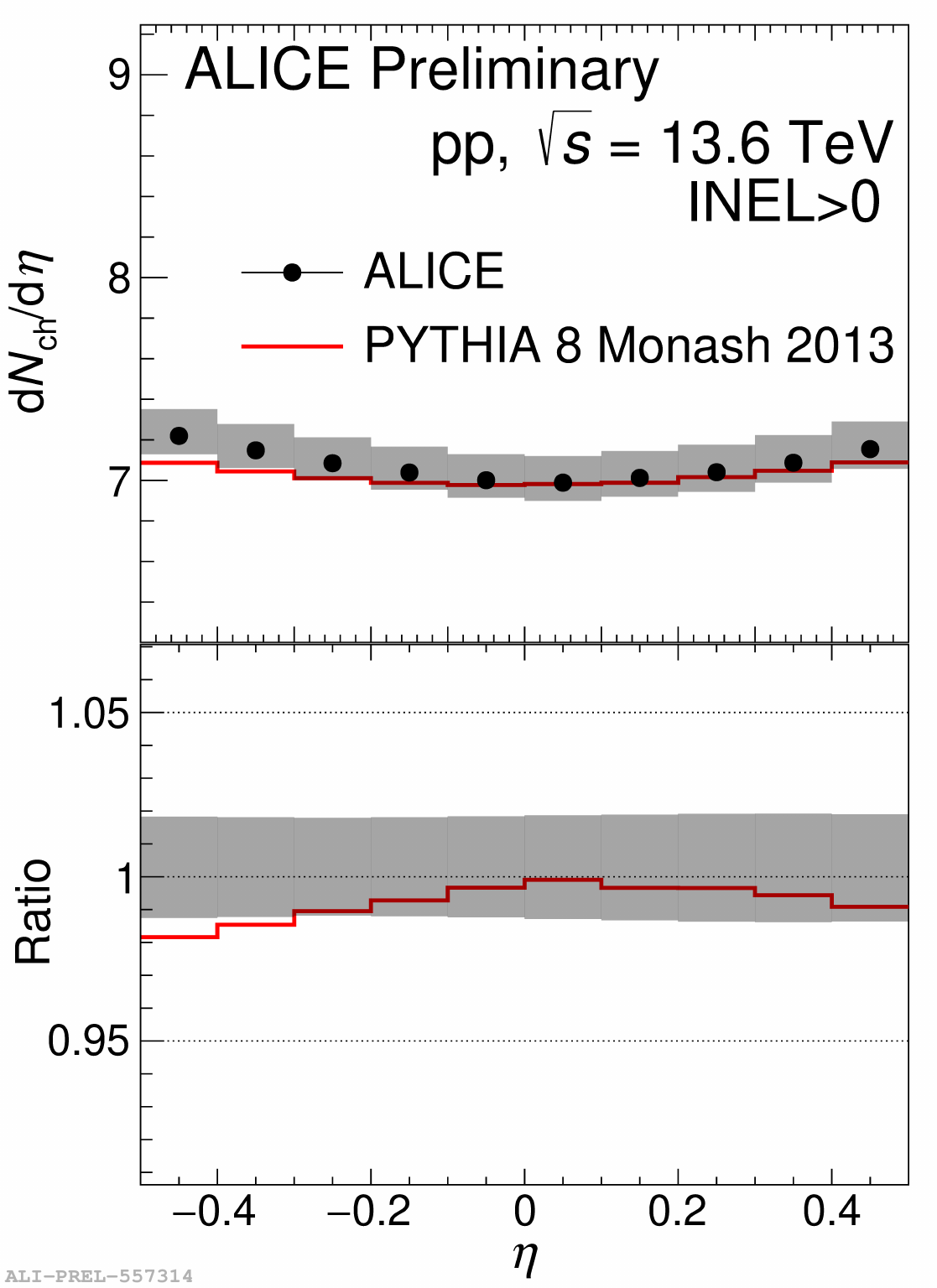}
  \end{minipage}\hfill
  \begin{minipage}[c]{0.58\textwidth}
    \includegraphics[scale=0.34]{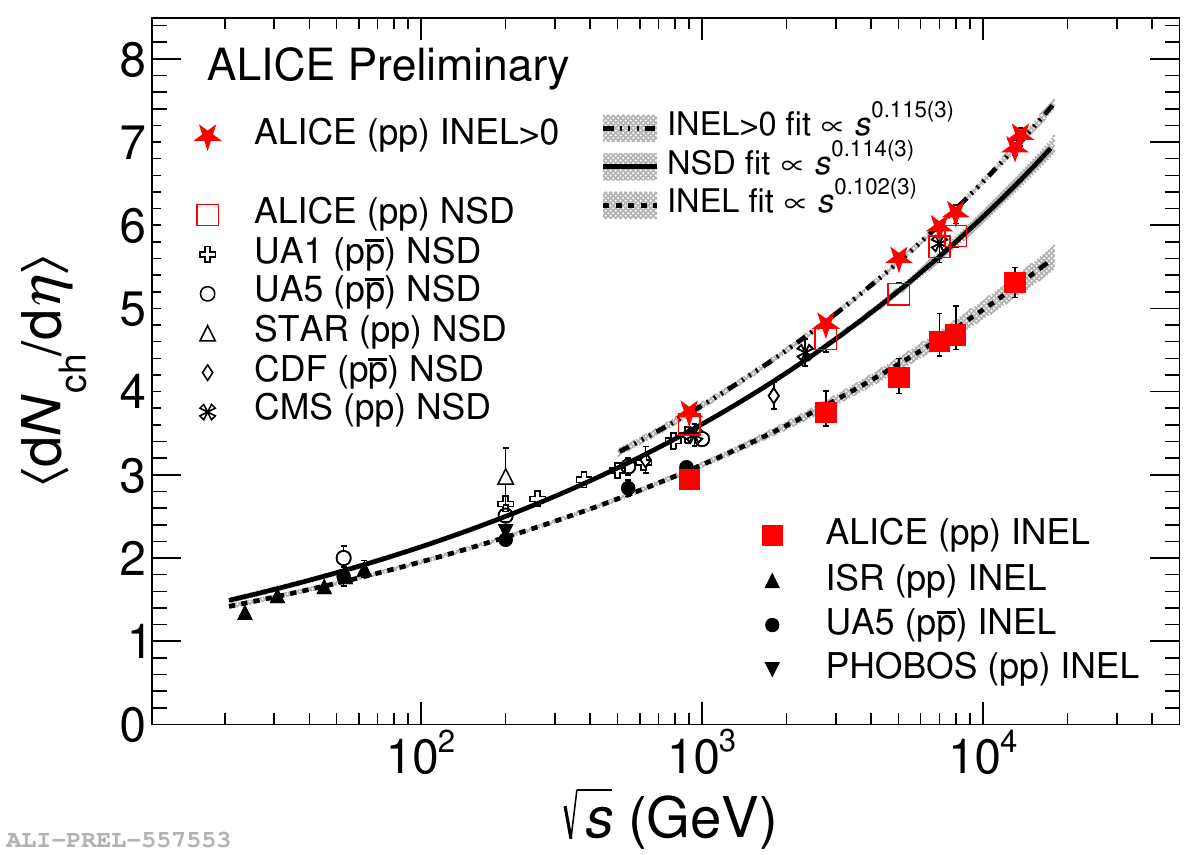}
    \caption{(Left) Distributions of \dndeta in \pp collisions at 13.6 TeV for INEL>0 event class. Data are compared with \textsc{Pythia}~8 (Monash 2013 tune) prediction. (Right) The energy dependence of \avdndeta in \pp collisions. Lines indicate fits with a power-law dependence on \s.}
    \label{dndeta_pp}
  \end{minipage}
\end{figure}

\section{Results and discussion}
\subsection{pp collisions at \s = 13.6 TeV}
The resulting pseudorapidity density distribution of charged particles in \pp collisions is presented in Fig.~\ref{dndeta_pp} (left). The distribution is measured for inelastic (INEL) \pp events having at least one charged particle produced in the interval $|\eta|<1$ (INEL$>$0 event class). The data are compared to prediction from MC event generator \textsc{Pythia}~8 with the Monash 2013 tune~\cite{Skands:2014pea}, showing that \textsc{Pythia}~8 reproduces the measurements well. In the central rapidity region ($|\eta|<0.5$), the \avdndeta is 7.12\,$^{+0.12}_{-0.09}$ which gives 10\% increase of charged-particle multiplicity as compared to 13~TeV with \avdndeta = 6.46\,$\pm$\,0.19~\cite{ALICE:pp13TeV}. Figure~\ref{dndeta_pp} (right) shows the energy dependence of \avdndeta in \pp collisions for INEL, non-single diffractive (NSD), and INEL$>$0 event classes. The lines are power-law fits of the energy dependence of the data for different event classes, and the gray bands represent the standard deviation of the fits. The \avdndeta at 13.6 TeV (solid red star symbol on extreme right) follows the established power-law trend as a function of \s.

\subsection{Pb--Pb collisions at \snn = 5.36 TeV}
\begin{figure}[h!]
  \centering
  \includegraphics[scale=0.34]{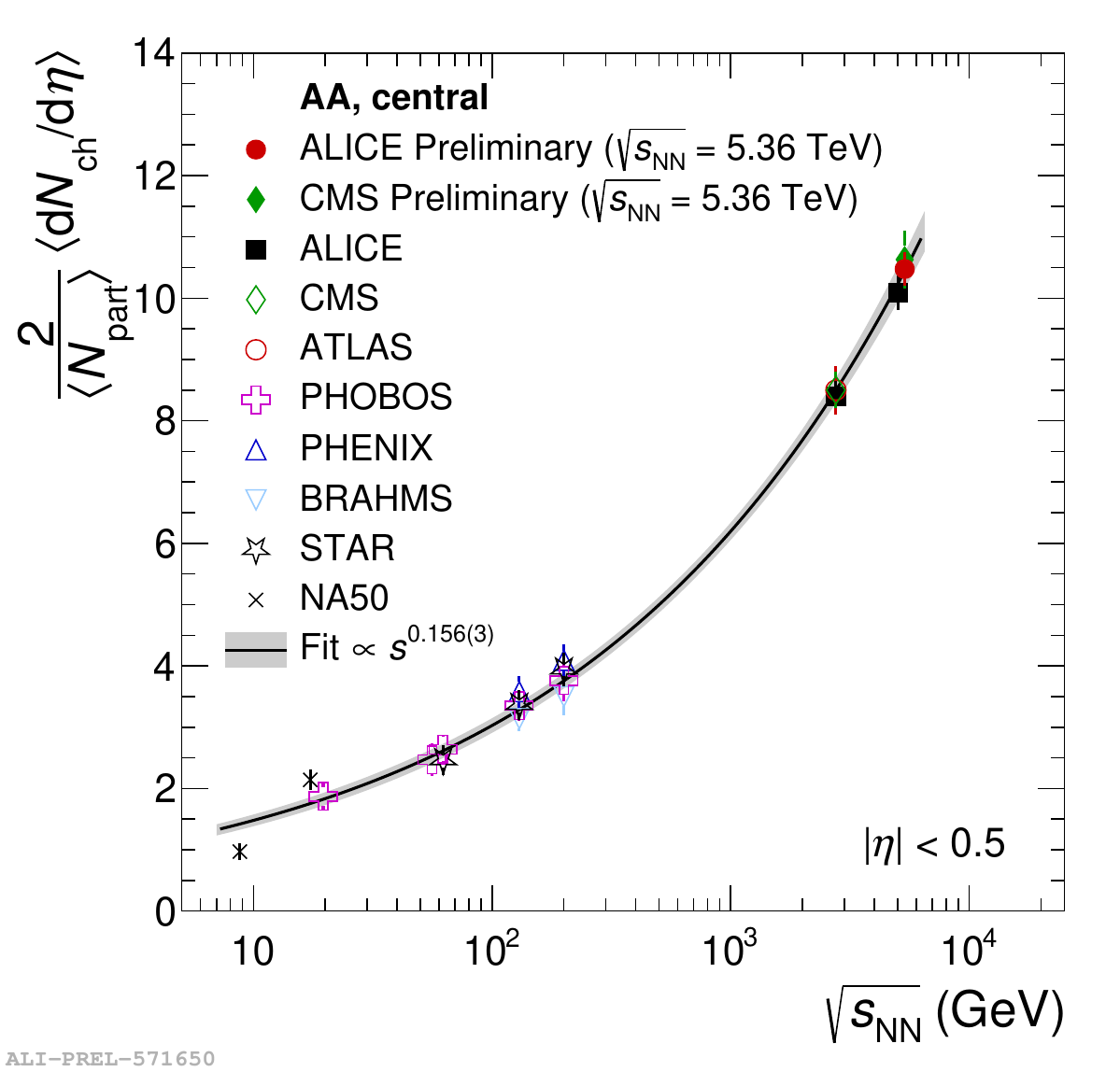}
  \includegraphics[scale=0.38]{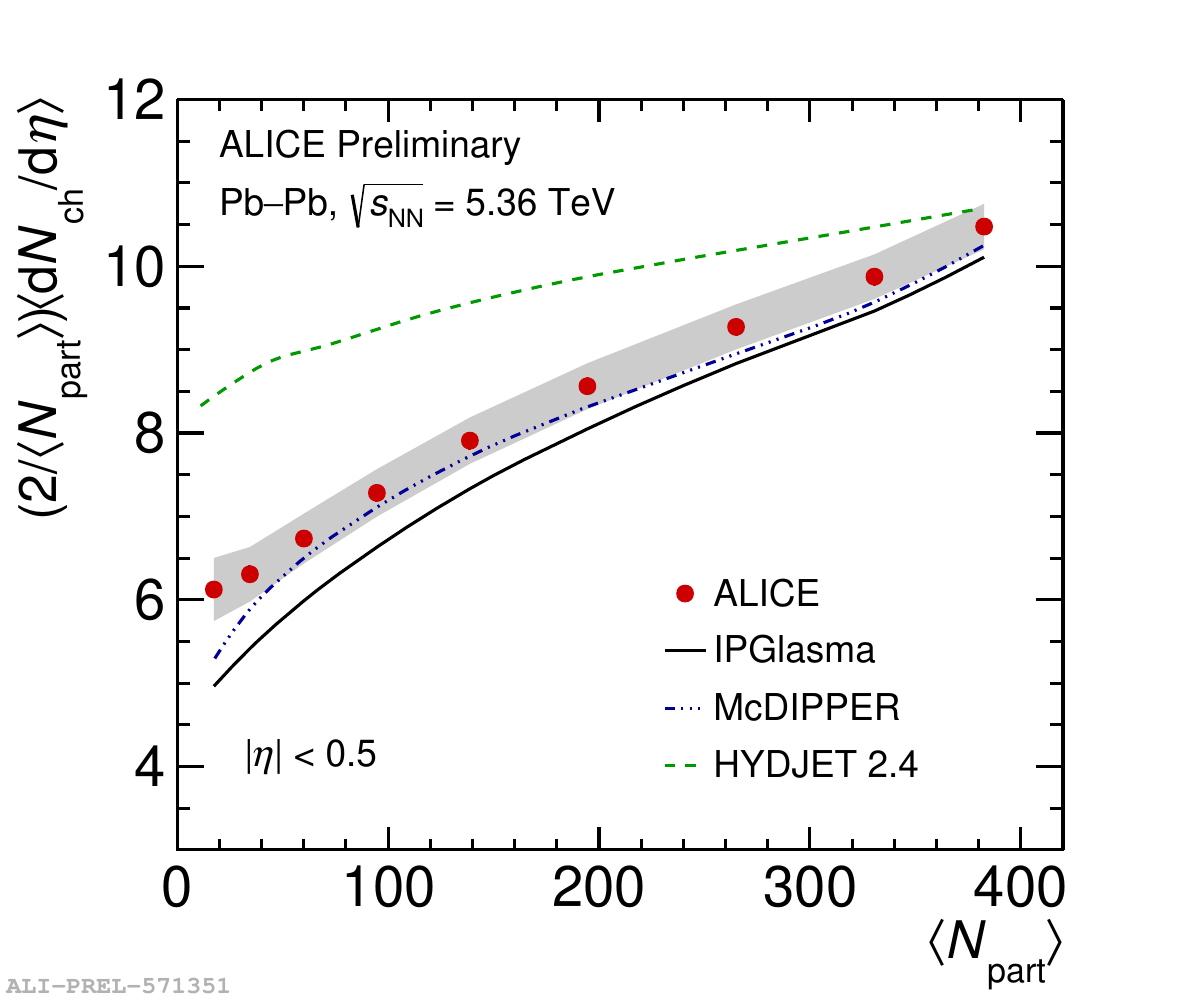}
  \caption{(Left) Values of (2/$\langle \npart \rangle$)\avdndeta for central heavy-ion collisions as a function of \snn. (Right) Centrality evolution of (2/$\langle \npart \rangle$)\avdndeta in \PbPb collisions at \snn = 5.36 TeV. Predictions from theoretical models are superimposed.}
  \label{dndetacent_PbPb}
\end{figure}

The \avdndeta over $|\eta|<0.5$ in \PbPb collisions is measured for nine centrality classes, ranging from 0--5\% (most central) to 70--80\% (most peripheral). To compare bulk particle production in different collision systems and at different energies, the charged-particle density is divided by the $\langle \npart \rangle/2$ determined for each centrality class. For the most central \PbPb collisions, the measured \avdndeta is 2004\,$\pm$\,52 and normalised per participant pair that corresponds to (2/$\langle \npart \rangle$)\avdndeta = 10.5\,$\pm$\,0.3. In Fig.~\ref{dndetacent_PbPb} (left), this value (solid red circle) is compared to the central heavy-ion collisions at lower centre-of-mass energies (see~\cite{ALICE:PbPbcent5p02} and references therein). The recent measurement at \snn = 5.36 TeV by CMS experiment~\cite{CMS:2024ykx} (solid green diamond) is also shown in the figure, which is consistent with our measurement. The dependence of (2/$\langle \npart \rangle$)\avdndeta on \snn is fitted with a power-law ($\alpha \cdot s_{\rm NN}^{\beta}$) that gives $\beta=0.156\pm0.003$. This $\beta$ value can be compared to $\beta=0.115\pm0.003$ obtained for \pp collisions (see Fig.~\ref{dndeta_pp} (right)) using the same power-law fit function. This shows that the charged-particle multiplicity increases faster with energy in central heavy-ion collisions compared to \pp collisions. The results at \snn =\,5.36 TeV confirms the trend established by lower-energy data since $\beta$ does not change significantly (earlier we found $\beta=0.155\pm0.004$~\cite{ALICE:PbPbcent5p02}) when the new point is included in the fit.

Figure~\ref{dndetacent_PbPb} (right) presents the measured (2/$\langle \npart \rangle$)\avdndeta (solid circles) as a function of $\langle \npart \rangle$ in \PbPb collisions at \snn =\,5.36 TeV. The gray bands represent the total systematic uncertainties, and the statistical uncertainties are within the symbol size. A strong centrality dependence is observed, with (2/$\langle \npart \rangle$)\avdndeta decreasing by a factor of $\simeq$1.7 from the most central collisions, large $\langle \npart \rangle$, to the most peripheral, small $\langle \npart \rangle$. This trend is similar to that previously measured in lower-energy \PbPb collisions~\cite{ALICE:PbPbcent2p76,ALICE:PbPbcent5p02}.

The centrality dependence of (2/$\langle \npart \rangle$)\avdndeta is also compared to predictions from HYDJET++ (v2.4) MC event generator and to hydrodynamic calculations using IP-Glasma and McDIPPER models. The HYDJET++~\cite{Lokhtin:2008xi} describes a heavy-ion collision as a superposition of the soft hydro-type component and the hard component resulting from multiparton fragmentation, treating both components independently. The hard component is described by PYTHIA simulation, while the soft component is modelled as a thermal hadronic state generated on the freeze-out hypersurfaces derived from the parametrisation of relativistic hydrodynamics. Hadron multiplicities are estimated using the effective thermal volume approximation. The IP-Glasma is an initial-state model~\cite{Schenke:IPglasma} based on the Color Glass Condensate (CGC) effective theory to describe the initial distribution of gluons in the colliding nuclei at different impact parameters, taking into account the phenomenon of gluon saturation. In this study, we consider a hybrid model~\cite{Schenke:IPglasma_music_urqmd} consisting of the IP-Glasma initial conditions followed by the MUSIC hydrodynamic model, coupled to a hadronic cascade model (UrQMD). The McDIPPER is also an initial-state model~\cite{Garcia-Montero:Mcdipper}, which calculates the energy and charge deposition in heavy-ion collisions based on the leading order cross-section calculations within the CGC framework. The value of \dndeta is estimated from the initial energy density profile using a hydrodynamical response as outlined in Ref.~\cite{Giacalone:2019ldn}. As shown in Fig.~\ref{dndetacent_PbPb} (right), both IP-Glasma+MUSIC+UrQMD (black line) and McDIPPER (blue line) calculations capture the general trend of the data; however, the McDIPPER model provides a better description of the data, particularly in peripheral \PbPb collisions, than the IP-Glasma+MUSIC+UrQMD model. HYDJET++ fails to reproduce both the shape and the magnitude of the dependence of multiplicity on centrality.

\section{Summary}
In summary, we have reported the measurements of charged-particle pseudorapidity density in \pp collisions at \s = 13.6 TeV and in \PbPb collisions at \snn = 5.36 TeV with the ALICE detector at the LHC. The meaured charged-particle densities in $|\eta|<0.5$ are 7.12\,$^{+0.12}_{-0.09}$ and 2004\,$\pm$\,52 for INEL$>0$ \pp collisions and most central \PbPb collisions, respectively. These quantities are in agreement with the expectations from lower energy extrapolations. The centrality dependence of \avdndeta at \snn = 5.36 TeV is found to be similar to that of previously measured lower-energy \PbPb collisions. Measurements of this fundamental observable of bulk particle production at new energy regimes can serve as essential inputs for developing theoretical models and MC event generators.


\bibliographystyle{utphys}   
\bibliography{bibliography}

\providecommand{\href}[2]{#2}\begingroup\raggedright\begin{thebibliography}{10}

\bibitem{ALICE:upgrade}
{\bfseries ALICE} Collaboration, S.~Acharya {\em et~al.}
  \href{http://dx.doi.org/10.1088/1748-0221/19/05/P05062}{{\em JINST}
  {\bfseries 19} no.~05, (2024) P05062}.

\bibitem{ALICE:ITSupgrade}
{\bfseries ALICE} Collaboration, B.~Abelev {\em et~al.}
  \href{http://dx.doi.org/10.1088/0954-3899/41/8/087002}{{\em J. Phys. G}
  {\bfseries 41} (2014) 087002}.

\bibitem{ALICE:TPCupgrade}
{\bfseries ALICE TPC} Collaboration, J.~Adolfsson {\em et~al.}
  \href{http://dx.doi.org/10.1088/1748-0221/16/03/P03022}{{\em JINST}
  {\bfseries 16} no.~03, (2021) P03022}.

\bibitem{ALICE:FIT}
{\bfseries ALICE} Collaboration, W.~H. Trzaska
  \href{http://dx.doi.org/10.1016/j.nima.2016.06.029}{{\em Nucl. Instrum. Meth.
  A} {\bfseries 845} (2017) 463--466}.

\bibitem{Miller:2007ri}
M.~L. Miller {\em et~al.}
  \href{http://dx.doi.org/10.1146/annurev.nucl.57.090506.123020}{{\em Ann. Rev.
  Nucl. Part. Sci.} {\bfseries 57} (2007) 205--243}.

\bibitem{ALICE:2013hur}
{\bfseries ALICE} Collaboration, B.~Abelev {\em et~al.}
  \href{http://dx.doi.org/10.1103/PhysRevC.88.044909}{{\em Phys. Rev. C}
  {\bfseries 88} no.~4, (2013) 044909}.

\bibitem{Bierlich:2018xfw}
C.~Bierlich, G.~Gustafson, L.~L\"onnblad, and H.~Shah
  \href{http://dx.doi.org/10.1007/JHEP10(2018)134}{{\em JHEP} {\bfseries 10}
  (2018) 134}.

\bibitem{ALICE:2015olq}
{\bfseries ALICE} Collaboration, J.~Adam {\em et~al.}
  \href{http://dx.doi.org/10.1140/epjc/s10052-016-4571-1}{{\em Eur. Phys. J. C}
  {\bfseries 77} no.~1, (2017) 33}.

\bibitem{Skands:2014pea}
P.~Skands, S.~Carrazza, and J.~Rojo
  \href{http://dx.doi.org/10.1140/epjc/s10052-014-3024-y}{{\em Eur. Phys. J. C}
  {\bfseries 74} no.~8, (2014) 3024}.

\bibitem{ALICE:pp13TeV}
{\bfseries ALICE} Collaboration, J.~Adam {\em et~al.}
  \href{http://dx.doi.org/10.1016/j.physletb.2015.12.030}{{\em Phys. Lett. B}
  {\bfseries 753} (2016) 319--329}.

\bibitem{ALICE:PbPbcent5p02}
{\bfseries ALICE} Collaboration, J.~Adam {\em et~al.}
  \href{http://dx.doi.org/10.1103/PhysRevLett.116.222302}{{\em Phys. Rev.
  Lett.} {\bfseries 116} no.~22, (2016) 222302}.

\bibitem{CMS:2024ykx}
{\bfseries CMS} Collaboration, A.~Hayrapetyan {\em et~al.}
  \href{http://arxiv.org/abs/2409.00838}{{\ttfamily arXiv:2409.00838
  [hep-ex]}}.

\bibitem{ALICE:PbPbcent2p76}
{\bfseries ALICE} Collaboration, K.~Aamodt {\em et~al.}
  \href{http://dx.doi.org/10.1103/PhysRevLett.106.032301}{{\em Phys. Rev.
  Lett.} {\bfseries 106} (2011) 032301}.

\bibitem{Lokhtin:2008xi}
I.~P. Lokhtin {\em et~al.}
  \href{http://dx.doi.org/10.1016/j.cpc.2008.11.015}{{\em Comput. Phys.
  Commun.} {\bfseries 180} (2009) 779--799}.

\bibitem{Schenke:IPglasma}
B.~Schenke, P.~Tribedy, and R.~Venugopalan
  \href{http://dx.doi.org/10.1103/PhysRevLett.108.252301}{{\em Phys. Rev.
  Lett.} {\bfseries 108} (2012) 252301}.

\bibitem{Schenke:IPglasma_music_urqmd}
B.~Schenke, C.~Shen, and D.~Teaney
  \href{http://dx.doi.org/10.1103/PhysRevC.102.034905}{{\em Phys. Rev. C}
  {\bfseries 102} no.~3, (2020) 034905}.

\bibitem{Garcia-Montero:Mcdipper}
O.~Garcia-Montero, H.~Elfner, and S.~Schlichting
  \href{http://dx.doi.org/10.1103/PhysRevC.109.044916}{{\em Phys. Rev. C}
  {\bfseries 109} no.~4, (2024) 044916}.

\bibitem{Giacalone:2019ldn}
G.~Giacalone {\em et~al.}
  \href{http://dx.doi.org/10.1103/PhysRevLett.123.262301}{{\em Phys. Rev.
  Lett.} {\bfseries 123} no.~26, (2019) 262301}.

\end{thebibliography}\endgroup

\end{document}